\documentclass[12pt,reqno,a4paper]{article}

\usepackage[british]{babel}
\usepackage[latin1]{inputenc}

\usepackage{amsmath}
\usepackage{amssymb}
\usepackage{amsthm}
\usepackage{latexsym}

\usepackage{xspace}

\newtheorem{thm}{Theorem}[section]
\newtheorem{lem}[thm]{Lemma}
\newtheorem{cor}[thm]{Corollary}
\numberwithin{equation}{section}

\newcommand{\ket}[1]{\left\vert #1 \right\rangle}
\newcommand{\n}[1]{\left\|#1\right\|}
\newcommand{\smap}[3]{#1 : #2 \longrightarrow #3\xspace}

\newlength{\strutspace}
\renewcommand{\strut}{\raisebox{-0.5\strutspace}{\rule{0ex}{\strutspace}}}

\title{A Note on the Quantum Query Complexity\\of the Hidden Subgroup Problem}
\author{Troels Windfeldt\footnote{The author has received financial support from Emil Herborgs Legat.}}

\date{}

\begin{document}

\maketitle


\begin{abstract}
We are concerned with the Hidden Subgroup Problem for finite groups. We present a simplified analysis of a quantum algorithm proposed
by Hallgren, Russell and Ta-Shma as well as a detailed proof of a
lower bound on the probability of success of the algorithm. 
\end{abstract}

\section{Introduction}
We are concerned with the following version of the Hidden Subgroup Problem. 

\textbf{Problem.}
\textit{Given a finite group $G$, a finite set $X$ and a map $\smap{f}{G}{X}$ which is constant on the left cosets of some unknown subgroup $H$ of $G$ and distinct on distinct cosets, determine the subgroup $H$.}

Shor's quantum algorithms for prime factorization and discrete logarithms \cite{shor}, Simon's quantum algorithm for the ``XOR-mask'' problem \cite{simon} as well as the open Graph Isomorphism Problem \cite{kobler} all reduce to instances of the Hidden Subgroup Problem (although in the case of prime factorization the group $G$ is not finite). In this note we study a quantum algorithm proposed to solve the Hidden Subgroup Problem in which the quantum Fourier transform has a significant role \cite{hallgren}.

We shall  require quantum registers capable of representing the elements of $G$ and $X$. Thus, let $\mathcal{H}_G$ and $\mathcal{H}_X$ denote quantum registers with orthonormal bases $\{\ket{g}\:|\: g\in G\}$ and $\{\ket{x}\:|\: x\in X\}$ indexed by the elements of $G$ and $X$, respectively. We suppose that the map $f$ is given as a unitary operator $\smap{U_f}{\mathcal{H}_G \otimes \mathcal{H}_X}{\mathcal{H}_G \otimes \mathcal{H}_X}$ such that
\[
\ket{g}\otimes\ket{x_0} \longmapsto \ket{g}\otimes\ket{f(g)}
\]
for some fixed $x_0\in X$. Next we briefly describe the quantum Fourier transform. A representation of $G$ is a homomorphism $\rho:G\longrightarrow U_{d_\rho}(\mathbb{C})$, where $U_{d_\rho}(\mathbb{C})$ denotes the group of unitary $d_\rho\times d_\rho$ matrices with complex entries. The set of all inequivalent irreducible representations is denoted $\widehat G$. Let $\mathcal{H}_{\widehat G}$ denote a quantum register with an orthonormal basis $\{\ket{\rho,i,j} \:|\: \rho\in \widehat G \text{ and } 1\leq i, j \leq d_\rho\}$ indexed by the elements of $\widehat G$ and their entries. The quantum Fourier transform $F:\mathcal{H}_G \longrightarrow \mathcal{H}_{\widehat G}$ is the unitary operator defined by
\[
\ket{g} \longmapsto \sum_{\substack{\rho\in \widehat G\\1\leq i, j\leq d_\rho}}\sqrt{\frac{d_\rho}{|G|}}\:\rho(g)_{ij}\ket{\rho,i,j}.
\]
The quantum experiment and algorithm we study are the following.

\textbf{Experiment.}
\textit{
\begin{itemize}
\item [\textit{1.}] Initialize a quantum system in the state $\ket{\rho_{\text{triv}}, 1, 1}\otimes\ket{x_0}$, where $\rho_{\text{triv}}$ denotes the trivial representation of $G$.
\strut\item [\textit{2.}] Apply the inverse quantum Fourier transform to the first register resulting in the uniform superposition
\[
\frac{1}{\sqrt{|G|}} \sum_{g\in G}\ket{g}\otimes\ket{x_0}.
\]
\strut\item [\textit{3.}] Apply $U_f$ resulting in the entangled state
\[
\frac{1}{\sqrt{|G|}} \sum_{c\in C}\sum_{h\in H}\ket{ch}\otimes\ket{f(c)},
\]
where $C$ denotes a complete set of coset representatives of the subgroup $H$ of $G$.
\strut\item [\textit{4.}] Apply the quantum Fourier transform to the first register resulting in the final state
\[
\frac{1}{|G|} \sum_{c\in C} \sum_{\substack{\rho \in \widehat G\\1\leq i, j \leq d_\rho}} \left(\sqrt{d_\rho}\sum_{h\in H}\rho(ch)_{ij}\right)\ket{\rho, i, j}\otimes\ket{f(c)}.
\]
\strut\item [\textit{5.}] Measure the first register and observe a basis vector $\ket{\rho, i, j}$.
\strut\item [\textit{6.}] Return the irreducible representation $\rho$.
\end{itemize}
}

\textbf{Algorithm.}
\textit{
\begin{itemize}
\item[\textit{1}.] Observe $n$ irreducible representations $\rho_1, \ldots, \rho_n$ by making independent trials of the experiment.
\strut\item[\textit{2}.]  Classically compute the intersection $N=\bigcap_{i=1}^n \ker \rho_i$ of the kernels of the irreducible representations.
\strut\item[\textit{3}.] Return the normal subgroup $N$.
\end{itemize}
}

In general it is not known how to implement the quantum Fourier
transform efficiently or how to calculate the intersection of the kernels of the irreducible representations that are measured. Thus we are interested in the query complexity of the algorithm which is the number of times it needs to evaluate the map $f$. Note that each trial of the experiment requires only one evaluation of $f$.

Obviously, the algorithm may return $H$ only if $H$ is a normal subgroup of $G$. In \cite{hallgren} it is shown that for $n=4\log_2|G|$ the algorithm returns the largest subgroup of $H$ that is normal in $G$ with high probability. Unfortunately, the proof presented there is somewhat unclear.

In this note we simplify the analysis of the probability distribution induced by the experiment. In particular, we completely avoid the entire discussion of both restricted and induced representations found in \cite{hallgren}. Furthermore, in the next section we give a lower bound on the probability of success of the algorithm as a function of $n$. More precisely, we prove the following theorem which has a curious corollary.

\begin{thm}\label{thm_hsp}
For $n>2\:\Omega(|G|)$ the algorithm returns the largest subgroup of $H$ that is normal in $G$ with probability at least
\[
1-1/\exp\left(\frac{8\left(\frac{n}{2}-\Omega(|G|)\right)^2}{9n}\right),
\]
where $\Omega(|G|)$ denotes the total number of prime factors of the order of $G$.
\end{thm}

For example, if we take $n=4\log_2|G|$ the algorithm succeeds with probability at least $1-1/\exp\left(\textstyle\frac{2}{9}\log_2|G|\right)$, which essentially is the statement of theorem 4.3 in \cite{hallgren}.

It follows from \cite[pp. 354--358]{hardy} that $\Omega(n)$ has normal order $\log\log n$. In particular, we have 
\begin{equation}\label{eqn_normal}
\Omega(n) \leq (1+\varepsilon)\log\log n
\end{equation}
for all $\varepsilon>0$ and almost all positive integers
$n$. \emph{Almost all} means that the fraction of positive integers
less than $x$ for which inequality \eqref{eqn_normal} hold tends to
$1$ as $x$ tends to infinity. Using this bound we obtain the following
corollary which in a sense says that in many cases the query
complexity is exponentially better than in the worst case where
$\Omega(|G|) = \log_2|G|$. 

\begin{cor}\label{thm_hsp_almost}
For all $\varepsilon>0$ and $n>2(1+\varepsilon)\log\log |G|$ the algorithm returns the largest subgroup of $H$ that is normal in $G$ with probability at least
\[
1-1/\exp\left(\frac{8\left(\frac{n}{2}-(1+\varepsilon)\log\log |G|\right)^2}{9n}\right),
\]
for almost all orders of $G$.
\end{cor}

For example, if we take $\varepsilon=3.5$ and $n=18\log\log |G|$ the algorithm succeeds with probability at least $1-1/\log|G|$ for almost all orders of $G$. We remark that with $\varepsilon=3.5$ direct calculations show that 99.92\% of all positive integers $n$ up to $10^9$ satisfy inequality \eqref{eqn_normal}.

\section{Analysis of the Algorithm}

The experiment induces a probability distribution on the set $\widehat G$. Let $X$ denote a random variable with this distribution. It is natural to ask, what the probability that $X$ equals a given irreducible representation is. An answer to this question is provided by the following lemma. Recall that the character $\chi_\rho:G\longrightarrow \mathbb{C}$ of a representation $\rho$ is defined by $\chi_\rho(g)=tr(\rho(g))$.

\begin{lem}\label{lem_prob_X}
If $\rho$ is an irreducible representation of $G$ then
\[
P(X = \rho) = \frac{d_\rho}{|G|}\sum_{h\in H}\chi_\rho(h).
\]
\end{lem}

\begin{proof}
By the definition of X, we have
\begin{eqnarray*}
P(X=\rho) & = & \frac{1}{|G|^2} \sum_{c\in C} \sum_{1 \leq i, j \leq
  d_\rho} \left\vert\sqrt{d_\rho}\sum_{h\in H}\rho(ch)_{ij}\right\vert^2 \\
& = & \frac{d_\rho}{|G|^2} \sum_{c\in C} \n{\sum_{h\in H}\rho(ch)}^2 \\
& = & \frac{d_\rho}{|G||H|}\n{\sum_{h\in H} \rho(h)}^2,
\end{eqnarray*}
as $\rho$ is a homomorphism and the Hilbert-Schmidt norm is unitarily invariant. Hence,
\begin{eqnarray*}
P(X = \rho) & = & \frac{d_\rho}{|G||H|}tr\left(\left(\sum_{h\in H}\rho(h)\right)^*\left(\sum_{h'\in H}\rho(h')\right)\right)\\
& = & \frac{d_\rho}{|G||H|}\sum_{h\in H}\sum_{h'\in H} tr\left(\rho(h)^{-1}\rho(h')\right) \\
& = & \frac{d_\rho}{|G||H|}\sum_{h\in H}\sum_{h'\in H} \chi_\rho\left(h^{-1}h'\right)\\
& = & \frac{d_\rho}{|G|}\sum_{h\in H} \chi_\rho(h),
\end{eqnarray*}
as $\rho(g)$ is unitary for all $g\in G$. 
\end{proof}

The second part of the algorithm suggests that instead of considering $X$ we should look at the transformed random variable $Y = \ker X$. As before, it is natural to ask, what the probability that $Y$ equals a given normal subgroup is. A partial answer to this question is provided by the following lemma. We will need the fact that for any finite group $G$ and any element $g\in G$, we have
\begin{equation}\label{eqn_d_chi}
\sum_{\rho\in \widehat G}d_\rho \chi_\rho(g) = \delta_e(g)|G|,
\end{equation}
where $e$ denotes the neutral element of $G$ and $\delta_e$ is the Kronecker delta.

\begin{lem}\label{lem_prob_Y}
If $N$ is a normal subgroup of $G$ then
\[
P(Y\supseteq N)=[N:N\cap H]^{-1}.
\]
\end{lem}

\begin{proof} 
By the definition of $Y$ and lemma \ref{lem_prob_X}, we have
\[
P(Y\supseteq N) = \frac{1}{|G|}\sum_{h\in H}\sum_{\substack{\rho\in \widehat{G}\\N\subseteq\ker\rho}}d_\rho\chi_\rho(h).
\]
If $\rho:G\longrightarrow U_{d_\rho}(\mathbb{C})$ is an irreducible representation of $G$ which is trivial on $N$ then the map $\widetilde{\rho}:G/N\longrightarrow U_{d_\rho}(\mathbb{C})$ defined by $\widetilde{\rho}(gN)=\rho(g)$ is a well-defined irreducible representation of the quotient group $G/N$. In this way, the irreducible representations which are trivial on $N$ correspond to the irreducible representations of $G/N$. It is clear that $d_\rho = d_{\widetilde{\rho}}$ and $\chi_\rho(g)=\chi_{\widetilde{\rho}}(gN)$ and therefore 
\begin{eqnarray*}
P(Y\supseteq N)&=&\frac{1}{|G|}\sum_{h\in H}\sum_{\widetilde{\rho}\in \widehat{G/N}}d_{\widetilde{\rho}}\chi_{\widetilde{\rho}}(hN) \\
&=& \frac{1}{|G|}\sum_{h\in H}\delta_N(hN)[G:N],
\end{eqnarray*}
where the last equality follows from equation \eqref{eqn_d_chi} when applied to the group $G/N$. But $hN=N$ if and only if $h\in N\cap H$ and so 
\[
P(Y\supseteq N) = \frac{1}{|G|}\sum_{h\in N\cap H}[G:N] = [N:N\cap H]^{-1}.
\]
This completes the proof.
\end{proof}

We note two simple consequences of lemma \ref{lem_prob_Y}. If $N\subseteq H$ then $[N:N\cap H]=1$ and so $P(Y\supseteq N)=1$. On the other hand, if $N\nsubseteq H$ then $[N:N\cap H]\geq2$ and so $P(Y\supseteq N)\leq \frac{1}{2}$. Thus, if we denote by $H^G$ the largest subgroup of $H$ that is normal in $G$ we may conclude that
\begin{equation}\label{eqn_1}
Y\supseteq H^G
\end{equation}
and
\begin{equation}\label{eqn_2}
0\leq P(Y\supseteq N)\cdot I_{\{N\nsubseteq H\}}\leq \textstyle\frac{1}{2}
\end{equation}
for any normal subgroup $N$ of $G$. With these two results, which are lemma 4.1 and 4.2 in \cite{hallgren}, we are in position to prove theorem \ref{thm_hsp}.

\begin{proof}
Let $X_1, \ldots, X_n$ denote independent random variables with the same distribution as $X$ and let $Y_i = \ker X_i$. The algorithm returns the largest subgroup of $H$ that is normal in $G$ with probability $1-P\left(Y_1 \cap \cdots \cap Y_n \neq H^G\right)$.

We will now construct what is sometimes known as a Doob type martingale to which we apply Azuma's inequality \eqref{eqn_azuma} in the appendix. Define indicator variables by $I_i = I_{\{Y_0\cap \cdots \cap Y_{i-1} \subseteq Y_{i}\}} \cdot I_{\{Y_0\cap \cdots \cap Y_{i-1} \nsubseteq H\}}$, where $Y_0$ is the random variable defined by $P(Y_0=G)=1$. Note that these indicators need not be independent. Still, the partial sums $Z_i = \sum_{j=1}^i (I_j - E(I_j \:\vert\: Y_0, \ldots, Y_{j-1}))$ of the random variables $I_i-E(I_i\:|\:Y_0,\ldots,Y_{i-1})$ constitute a martingale. The fact that the sequence $Z_1, \ldots, Z_n$ is indeed a martingale with mean $Z_0=0$ follows from lemma \ref{lem_mar} in the appendix.

Suppose $Y_1 \cap \cdots \cap Y_n \neq H^G$ and consider the descending chain of normal subgroups
\[
Y_0 \supseteq Y_0\cap Y_1 \supseteq Y_0\cap Y_1 \cap Y_2 \supseteq \cdots \supseteq Y_0\cap \cdots \cap Y_n.
\]
Now suppose, indirectly, that $Y_0 \cap \cdots \cap Y_{i-1} \subseteq H$ for some $i$. This clearly implies that $Y_1 \cap \cdots \cap Y_n \subseteq H$ and so $Y_1 \cap \cdots \cap Y_n \subseteq H^G$, as $H^G$ is the largest subgroup of $H$ that is normal in $G$. However, by equation \eqref{eqn_1} we also have the reverse inclusion which is a contradiction. Thus, we must have $Y_0\cap \cdots \cap Y_{i-1} \nsubseteq H$ and therefore $I_i = I_{\{Y_0\cap \cdots \cap Y_{i-1} \subseteq Y_{i}\}} = I_{\{Y_0\cap \cdots \cap Y_{i-1} = Y_0\cap \cdots \cap Y_i\}}$. That is, the sum $\sum_{i=1}^n (1-I_i)$ counts the number of strict inclusions in the above chain. This number is necessarily less than or equal to $\Omega(|G|)$ and therefore $\sum_{i=1}^n I_i \geq n - \Omega(|G|)$. Thus, we see that
\begin{align*}
P\left(Y_1 \cap \cdots \cap Y_n \neq H^G\right) &\leq P\left(\sum_{i=1}^nI_i \geq n -\Omega(|G|) \right) \\&= P\left(Z_n\geq n-\Omega(|G|)-\sum_{i=1}^nE(I_i\:|\:Y_0, \ldots, Y_{i-1})\right)\\&\leq P\left(Z_n\geq \frac{n}{2}-\Omega(|G|)\right),
\end{align*}
where the last inequality follows from lemma \ref{lem_cond_exp} in the appendix. 

We are now almost in position to apply Azuma's inequality \eqref{eqn_azuma}. As $\frac{n}{2}-\Omega(|G|)>0$ we only need to verify that the martingale $Z_1, \ldots, Z_n$ has bounded differences $Z_i-Z_{i-1} = I_i-E(I_i\:|\:Y_0,\ldots, Y_{i-1})$. This follows from
\[
-\textstyle\frac{1}{2}\leq I_i-\textstyle\frac{1}{2}\leq I_i - E(I_i\:|\:Y_0, \ldots, Y_{i-1})\leq I_i\leq 1,
\]
where we have used lemma \ref{lem_cond_exp}, and so
\[
P\left(Y_1 \cap \cdots \cap Y_n \neq H^G\right) \leq \exp\left(-\frac{2\left(\frac{n}{2}-\Omega(|G|)\right)^2}{n\left(\frac{1}{2}+1\right)^2}\right).
\]
This completes the proof.
\end{proof}

It should be mentioned that there is a simpler argument, which also utilizes \eqref{eqn_1} and \eqref{eqn_2}, for the fact that for $n\approx\log_2^2|G|$ the algorithm returns the largest subgroup of $H$ that is normal in $G$ with high probability.\footnote{Private communication with Alexander Russell.} Unfortunately, it is not clear how to obtain corollary \ref{thm_hsp_almost} from this simpler argument.

\appendix
\section{Probability Theory}

The conditional expectation $E(X\:|\:Y)$ of a real discrete random variable $X$ given any discrete random variable $Y$ is defined whenever $P(Y=y)>0$ as the random variable which takes the value
\[
E(X\:|\:Y=y) = \sum_{x}xP(X=x\:|\:Y=y)
\]
with probability $P(Y=y)$, where the sum is over all the outcomes of $X$. We define the conditional expectation $E(X\:|\:Y_1, \ldots, Y_m)$ of $X$ given any discrete random variables $Y_1, \ldots, Y_m$ similarly.

A sequence $Z_1, \ldots, Z_n$ of real discrete random variables is said to be a martingale if
\[
E(Z_{i+1}\:|\:Z_1, \ldots, Z_i) = Z_i
\]
for all $i$. Taking expectations and applying equation \eqref{eqn_ce_1} below we see that $E(Z_i)=E(Z_1)$ for all $i$. Thus, it makes sense to speak about the mean of a martingale.

Suppose $Z_1, \ldots, Z_n$ is a martingale with mean $Z_0 = E(Z_1)$ and bounded differences. That is, for some non-negative constants $\alpha$ and $\beta$ we have $-\alpha\leq Z_i - Z_{i-1}\leq\beta$ for all $i$. Azuma's inequality then says that for all $i$ and $a>0$
\begin{equation}\label{eqn_azuma}
P(Z_i-Z_0\geq a) \leq \exp\left(-\frac{2a^2}{i(\alpha+\beta)^2}\right).
\end{equation}
The reader is referred to \cite[p. 307--308]{ross} from which this version of Azuma's inequality has been adapted.

In the following two lemmas we use the notation from the proof of theorem \ref{thm_hsp}. We will need that for any discrete random variables $Y_1, \ldots, Y_m$ and $Z_1, \ldots, Z_n$ the following properties hold.
\begin{align}\label{eqn_ce_1}
E(E(X\:|\:Y_1, \ldots, Y_m))&=E(X),\\\label{eqn_ce_2}
E(E(X\:|\:Y_1, \ldots, Y_m,Z_1, \ldots, Z_n)\:|\:Z_1, \ldots, Z_n)&=E(X\:|\:Z_1, \ldots, Z_n),\\\label{eqn_ce_3}
E(X\:|\:Y_1, \ldots, Y_m)&=X \\\label{eqn_ce_4}\nonumber
& \text{if $X$ is a function of $Y_1, \ldots, Y_m$,}\\
E(X\:|\:Y_1, \ldots, Y_m,Z_1, \ldots, Z_n)&=E(X\:|\:Y_1, \ldots, Y_m)\\\nonumber
& \text{if $Z_i$ is a function of $Y_1, \ldots, Y_m$.}
\end{align}

The proofs of these properties may be found in any standard reference on stochastic calculus. See for example \cite{mikosch} and \cite{ross}.

\begin{lem}\label{lem_mar}
The sequence $Z_1, \ldots, Z_n$ is a martingale with mean $Z_0 = 0$.
\end{lem}

\begin{proof}
Note that $Z_{i+1} = Z_i+I_{i+1}-E(I_{i+1}\:|\:Y_0, \ldots, Y_i)$ and so by linearity of the conditional expectation
\begin{multline*}
E(Z_{i+1}\:|\:Z_1, \ldots, Z_i) \\= E(Z_i\:|\:Z_1, \ldots, Z_i) +E(I_{i+1}\:|\:Z_1, \ldots, Z_i)-E(E(I_{i+1}\:|\:Y_0, \ldots, Y_i)\:|\:Z_1, \ldots, Z_i).
\end{multline*}
Now, $E(Z_i\:|\:Z_1, \ldots, Z_i)=Z_i$ by equation \eqref{eqn_ce_3} and
\begin{multline*}
E(E(I_{i+1}\:|\:Y_0, \ldots, Y_i)\:|\:Z_1, \ldots, Z_i) \\= E(E(I_{i+1}\:|\:Y_0, \ldots, Y_i, Z_1, \ldots, Z_i)\:|\:Z_1, \ldots, Z_i) \\=E(I_{i+1}\:|\:Z_1, \ldots, Z_i)
\end{multline*}
by equations \eqref{eqn_ce_4} and \eqref{eqn_ce_2}. Thus, we see that $E(Z_{i+1}\:|\:Z_1, \ldots, Z_i)=Z_i$ which proves that $Z_1, \ldots, Z_n$ is a martingale. It has mean
\[
Z_0 = E(Z_1) = E(I_1) - E(E(I_1\:|\:Y_0))=E(I_1)-E(I_1)=0
\]
by equation \eqref{eqn_ce_1}.
\end{proof}

\begin{lem}\label{lem_cond_exp}
We have $0 \leq E(I_i \:\vert\: Y_0, \ldots, Y_{i-1}) \leq \frac{1}{2}$.
\end{lem}

\begin{proof}
By the definition of conditional expectation
\[
E(I_i \:\vert\: Y_0=N_0, \ldots, Y_{i-1}=N_{i-1}) = P(I_i = 1 \:\vert\: Y_0=N_0, \ldots, Y_{i-1}=N_{i-1})
\] for all normal subgroups $N_0, \ldots, N_{i-1}$ of $G$ for which $P(Y_0=N_0, \ldots, Y_{i-1}=N_{i-1}) > 0$. Therefore,
\begin{multline*}
E(I_i \:\vert\: Y_0=N_0, \ldots, Y_{i-1}=N_{i-1}) \\
= \frac{P(Y_0\cap \cdots \cap Y_{i-1} \subseteq Y_{i}, Y_0\cap \cdots \cap Y_{i-1} \nsubseteq H, Y_0=N_0, \ldots, Y_{i-1}=N_{i-1})}{P(Y_0=N_0, \ldots, Y_{i-1}=N_{i-1})}\\
= \frac{P(N \subseteq Y_{i}, Y_0=N_0, \ldots, Y_{i-1}=N_{i-1}) \cdot I_{\{N \nsubseteq H\}}}{P(Y_0=N_0, \ldots, Y_{i-1}=N_{i-1})},
\end{multline*}
where $N = N_0\cap \cdots \cap N_{i-1}$ for short. As the random variables $Y_0, \ldots, Y_i$ are independent
\begin{multline*}
E(I_i \:\vert\: Y_0=N_0, \ldots, Y_{i-1}=N_{i-1}) \\
= \frac{P(N \subseteq Y_{i})P(Y_0=N_0, \ldots, Y_{i-1}=N_{i-1})\cdot I_{\{N \nsubseteq H\}}}{P(Y_0=N_0, \ldots, Y_{i-1}=N_{i-1})}
\end{multline*}
and the result follows by equation \eqref{eqn_2}.
\end{proof}

\section*{Acknowledgement}

I thank Sean Hallgren, Alexander Russell, Henrik Schlichtkrull and Jan Philip Solovej for discussions. I also thank Martin Jacobsen and Thomas Mikosch for many helpful discussions on probability theory.

\footnotesize{Department of Mathematics, Universitetsparken 5, 2100
  Copenhagen Ø, Denmark; e-mail: windfeldt@math.ku.dk}

\end{document}